\documentclass[sigconf,authorversion,natbib=true]{acmart}

\AtBeginDocument{%
  }

\acmConference[ISSTA Companion '25]{34th ACM SIGSOFT International
Symposium on Software Testing and Analysis}{June 25--28, 2025}{Trondheim,Norway}
\acmBooktitle{34th ACM SIGSOFT International Symposium on Software Testing and Analysis (ISSTA Companion '25), June 25--28, 2025, Trondheim, Norway}
\acmDOI{10.1145/3713081.3731728}
\acmISBN{979-8-4007-1474-0/2025/06}

\usepackage{soul,xcolor}
\usepackage[acronym]{glossaries}
\usepackage{multirow}
\usepackage[most]{tcolorbox} 
\usepackage{listings}
\usepackage{booktabs}
\usepackage{makecell}
\usepackage{hyperref}
\usepackage{subfigure}
\usepackage{subcaption}
\usepackage{graphicx}
\usepackage{multicol}
\usepackage{float}
\usepackage{pgfplots}
\usepackage{pgfplotstable}
\pgfplotsset{compat=1.17}
\usepackage{adjustbox}
\usepackage{array}
\usepackage{pgfplots}
\usepackage{caption}
\pgfplotsset{compat=1.18} 

\newcommand{\name}{ReGraph}

\newacronym{isa}{ISA}{Instruction Set Architecture}
\newacronym{plc}{PLC}{Programmable Logic Controllers}
\newacronym{iot}{IOT}{Internet Of Things}
\newacronym{cfg}{CFG}{Control Flow Graph}
\newacronym{acfg}{ACFG}{Attributed Control Flow Graph}
\newacronym{ot}{OT}{Operational Technology}
\newacronym{gnn}{GNN}{Graph Neural Network}
\newacronym{cpg}{CPG}{Code Property Graph}
\newacronym{nlp}{NLP}{Natural Language Processing}
\newacronym{ir}{IR}{Intermediate Representation}
\newacronym{gat}{GAT}{Graph Attention Network}
\newacronym{ics}{ICS}{Industrial Control System}
\newacronym{ast}{AST}{Abstract Syntax Tree}
\newacronym{gmn}{GMN}{Graph Matching Network}
\newacronym{cots}{COTS}{Commercial Off-The-Shelf}

\definecolor{SpringGreen}{RGB}{198,220,103}
\begin{document}

\title{\name: A Tool for Binary Similarity Identification}


\author{Li Zhou}
\affiliation{%
  \institution{KAUST}
  \country{KSA}}
\email{li.zhou@kaust.edu.sa}

\author{Marc Dacier}
\affiliation{%
  \institution{KAUST}
  \country{KSA}}
\email{marc.dacier@kaust.edu.sa}

\author{Charalambos Konstantinou}
\affiliation{%
  \institution{KAUST}
  \country{KSA}}
\email{charalambos.konstantinou@kaust.edu.sa}

\renewcommand{\shortauthors}{Li \textit{et al.}}


\begin{abstract}

  Binary Code Similarity Detection (BCSD) is not only essential for security tasks such as vulnerability identification but also for code copying detection, yet it remains challenging due to binary stripping and diverse compilation environments.
  Existing methods tend to adopt increasingly complex neural networks for better accuracy performance. The computation time increases with the complexity. Even with powerful GPUs, the treatment of large-scale software becomes time-consuming. 
  To address these issues, we present a framework called~\name~to efficiently compare binary code functions across architectures and optimization levels. 
  Our evaluation with public datasets highlights that \name~exhibits a significant speed advantage, performing 700 times faster than \acrfull{nlp}-based methods while maintaining comparable accuracy results with respect to the state-of-the-art models.
  \end{abstract}




\maketitle

\section{Introduction}

Binary Code Similarity Detection (BCSD) plays a critical role in various downstream applications such as vulnerability identification, firmware security analysis, and software reuse detection~\cite{basit2005detecting, luo2014semantics, zhao2022large, shirani2018b, huang2017binsequence}. Despite significant progress, it remains an active research area. A key difficulty lies in that, even when compiled from the same source code, binaries can exhibit substantial differences across various compilation options in both structure and semantics. Furthermore, to minimize file size, binary stripping is commonly employed~\cite{marcelli2022machine}, eliminating symbol tables and debug information~\cite{GNU_strip}, thereby exacerbating the challenge of binary-level code analysis~\cite{wang2024binenhance}. 

Manually reverse engineering binary code functions is a solution but a very time-consuming one, especially when analyzing millions of stripped binary functions~\cite{xu2021innovative}. Hence, research directions focus on automatic methods. At first, methods have compared binary file elements such as operator counts, jumps, and Control Flow Graphs (CFGs)~\cite{xorpd, david2014tracelet}, but different compilation options significantly alter \acrshort{cfg}s and other extracted features~\cite{mengin2021binary}, limiting their performance in cross-optimization and cross-architecture scenarios~\cite{pewny2015cross,marcelli2022machine}. More recently, new methods have been proposed to leverage neural networks and have outperformed earlier methods~\cite{xu2017neural,yu2020codecmr,yu2020order}. Despite that, the differences caused by different compilation options require an increasing amount of neural network parameters~\cite{chandramohan2016bingo,hu2016cross, xu2017neural}, as well as the creation of large models~\cite{ding2019asm2vec,luo2023vulhawk,yu2020codecmr}. These models require powerful and expensive GPUs and render the computation time impractical when analyzing large amounts of functions.

To address these challenges, we propose \name, a function-level cross-architecture and cross-compilation binary code matching framework. \name~efficiently identifies known functions within a given software by leveraging code lifting, re-optimization, Code Property Graphs (CPGs), and Graph Neural Networks (GNNs) to compute function similarity. This enables analysts to further examine the most likely similar functions within a binary executable. We evaluate \name~on public datasets~\cite{marcelli2022machine}. The results demonstrate that our framework exhibits high accuracy and fast matching speed with much lower resource consumption than state-of-the-art solutions. Our contributions can be summarized as follows: 

\begin{itemize}

    \item We propose mitigating binary discrepancies by offloading them to the lifter and re-optimizer. Our experiments show that these techniques improve similarity scores by an average of 72.8\% with respect to those reported by BinDiff~\cite{Bindiff_2011}. Our results demonstrate that these techniques can serve as a generalizable meta-method for other binary code similarity detection frameworks.
    
    \item We integrate \acrshort{cpg}s and \acrshort{gnn}s with lifting and re-optimization to develop~\name. \name~achieves a 700× speedup over NLP-based methods while maintaining comparable performance to state-of-the-art approaches.

    \item With \name, our proposed command-line tool, analysts can focus on the top-K most similar functions rather than analyzing the entire binary file, significantly reducing their workload.

\end{itemize}

This paper is structured as follows: Section II presents the usage of our tool, and Section III provides a detailed explanation of each component of \name. We evaluate \name~in Section IV. Section V concludes the paper and discusses potential directions for future research.

\section{Usage}
In this section, we demonstrate the usage of \name, A Python-based command-line tool that detects similar binary code functions. It operates in two phases: the training phase, which processes newly encountered binary files, and the inference phase, which computes similarity results based on the trained model.

\subsection{Requirements}

\name~is built in Python with additional dependencies. It requires RetDec~\cite{kvroustek2017retdec} for binary lifting, LLVM~\cite{lattner2002llvm} for re-optimization, and Joern~\cite{Yamaguchi_2014} for extracting CPGs. These dependencies should be installed before using \name.  

\subsection{Training}

\textbf{Preprocessing:} Preprocessing includes binary lifting, re-optimization, CPG extraction, and CPG vectorization. We have integrated the first three functionalities into a Python script, allowing them to be executed with a single command. By default, we assume the directory structure follows the format \texttt{project/architecture/
optimization-level}. If not, users can specify custom file paths through the extra-provided Python programming interface. Once CPGs are extracted, a separate command-line script vectorize the CPGs into datasets for training and inference. Additionally, an auxiliary file, the \texttt{op\_file}, will be generated to store statistics about operators extracted from binaries to assist in encoding the CPGs.  

\textbf{Training:} To achieve optimal performance on new binary files, we recommend training the model on these files. All hyperparameters governing the training process are defined within the configuration file called \texttt{train\_config.yaml}. Subsequent to data preprocessing, this file requires modification to specify the dataset directory path and finalize the parameters for training execution. Then, running the model training script will initiate the training process and generate the trained model file in the specified output folder.

\subsection{Inference}
With the trained model file and the corresponding \texttt{op\_file}, inference can be performed. Given a target function in a binary file (target binary file), our tool identifies the top K's most similar functions, in which the user pre-defines the K, from another binary file (candidate binary file). To minimize user interaction, our tool outputs all functions from the target binary file along with their top-K matches from the candidate binary file, including the corresponding similarity scores, rather than requiring users to select a target function manually. This functionality is encapsulated in a script, where users specify the pre-trained model path, the \texttt{op\_file}, and the two binaries via the command line to generate the results in the \texttt{Excel} spreadsheet.

\subsection{Illustrative Example}
In this section, we present an example using two files from OpenPLC (a collection of open-source programs for Programmable Logic Controllers, PLCs)~\cite{alves2014openplc}, compiled under different environments. The target binary is compiled for the x86 architecture with \texttt{-O0} optimization, while the candidate binary is compiled for the ARM architecture with \texttt{-O3} optimization. We use the function \texttt{\_\_time\_sub} as a case study. Despite originating from the same source code,  their corresponding assembly code demonstrates significant differences, as shown in Figure~\ref{fig:asm_comparison}.

\begin{figure}[ht]
    \centering
    \begin{minipage}{0.48\linewidth}
        \begin{lstlisting}
MOV R12, R0
SUB SP, SP, #8
PUSH {R4, LR}
...
ADD SP, SP, #8
BX LR
        \end{lstlisting}
    \end{minipage}
    \hfill
    \begin{minipage}{0.48\linewidth}
        \begin{lstlisting}
push ebp
mov ebp, esp
sub esp, 28h
...
leave
retn 4
        \end{lstlisting}
    \end{minipage}
    \vspace{-9mm}
    \caption{The assembly code of example functions compiled in ARM -O3 (left) and X86 -O0 (right).}
    \vspace{-3mm}
    \label{fig:asm_comparison}
\end{figure}

After training \name~and utilizing the pre-trained model, we set K = 5. Table~\ref{tab:topk_functions} presents the example output of functions similar to \texttt{\_\_time\_sub}, In stripped binaries, function names are typically removed. In such cases, security analysts can inspect the top-K functions to identify the target function, rather than searching through the entire binary, which significantly reduces the workload and improves efficiency in binary analysis. To aid in understanding the results, we provide the function names in the last column, labeled \textit{Real Name}. Due to minor operator differences between \texttt{\_\_time\_sub} and \texttt{\_\_time\_add}, their similarity scores are close. In contrast, \texttt{INTEGRAL\_body\_\_} exhibits more structural differences from \texttt{\_\_time\_sub}, resulting in a lower similarity score.

\begin{table}[ht]
    \centering
    \begin{tabular}{cccc}
        \toprule
        \textbf{Top K} & \textbf{Score} & \textbf{Stripped Name} & \textit{\textbf{Real Name}} \\
        \midrule
        1 & 0.921 & function\_154  & \textit{\texttt{\_\_time\_sub}} \\
        2 & 0.917 & function\_104  & \textit{\texttt{\_\_time\_add}} \\
        3 & 0.809 & function\_2cc8 & \textit{\texttt{INTEGRAL\_body\_\_}} \\
        4 & 0.806 & function\_35b8 & \textit{\texttt{PROG0\_body\_\_}} \\
        5 & 0.797 & function\_74   & \textit{\texttt{\_\_normalize\_timespec}} \\
        \bottomrule
    \end{tabular}
    \caption{Top K functions with corresponding scores and addresses.}
    \label{tab:topk_functions}
\end{table}

\vspace{-6mm}

\section{\name}

\begin{figure*}[htbp]
    \centering
    \includegraphics[width=0.8\linewidth]{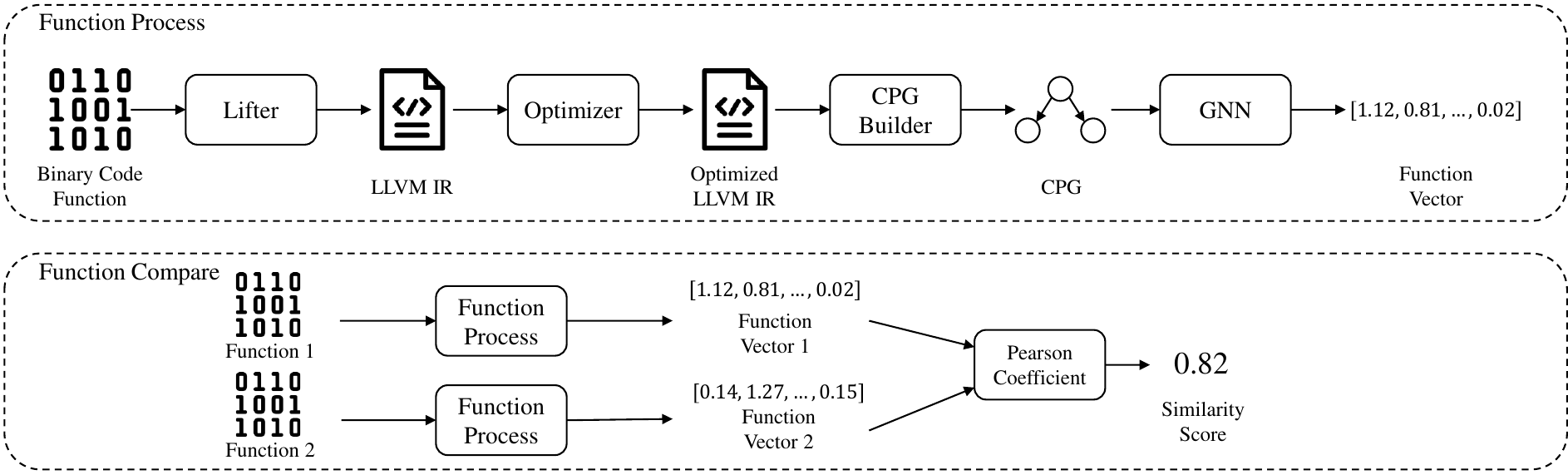}
    \caption{The structure of~\name.}
    \label{fig:framework}
    \Description{The structure of~\name.}
\end{figure*}

In this section, we detail each component of \name, whose\ structure is depicted in Figure~\ref{fig:framework}. Given a target binary code function, our objective is to measure its similarity score with each function of a test set via their corresponding \acrshort{cpg}s. \name~first uses a lifter to lift the binary code function to LLVM Intermediate Representation~(IR)~\cite{lattner2002llvm}. Then, the optimizer further re-optimizes it to the -O3 level, representing the highest optimization level commonly used in compilation. 

However, when the lifter and optimizer attempt to reorganize the binary code, in some cases, differences persist. Hence, we employ \acrshort{cpg} and \acrshort{gnn} to tolerate these differences. Once we obtain the optimized LLVM IR, we decompile it into pseudo-C code and convert it into a CPG using Joern~\cite{Yamaguchi_2014}. We then encode the CPG into the \acrshort{gnn} to obtain its corresponding function vector that represents the whole function. The similarity score is finally computed using the \textit{Pearson coefficient}~\cite{stat462_correlation} applied to their respective function vectors. A higher coefficient score indicates a greater similarity between the two functions. In the rest of this section, we detail the design choices for each component.


\subsection{Lifter}

Decompilers such as IDA Pro~\cite{Hex_1996} and Ghidra~\cite{Agency_2019} are widely used in reverse engineering. These decompilers maintain their own IR systems as fundamental platforms supporting their decompilation analysis frameworks, which lift all binary files to intermediate levels where subsequent optimizations and analyses are performed. The IR system decouples program logic analysis from specific CPU architectures, eliminating the need to implement program analysis and optimization methods for each architecture.  However, the IR generated by decompilers is primarily intended to produce more readable pseudo-C code rather than implement the optimizations typically found in compilers. Therefore, we choose to use another lifter, RecDec~\cite{kvroustek2017retdec}, that lifts binary codes from diverse architectures into standardized LLVM IR, thereby reducing differences caused by varying architectures.

\subsection{Re-optimization}

Different software vendors employ different optimization options to meet their specific requirements. According to the documentation of GCC~\cite{GNU_2023} and Clang~\cite{Clang_2024}, the relationship between different optimization levels is hierarchical. For instance, -O2 includes all optimization options from -O1 and adds its own additional optimizations. -O3 encompasses all optimization options from -O2 and represents the highest optimization level. Applying -O3 to code that has already been optimized at lower levels optimizes the code to its maximum potential. Conversely, applying lower optimization levels to code undergoing -O3 optimization does not yield additional changes, as -O3 has already implemented all available optimizations. Therefore, if we can re-optimize the program to the highest optimization level, the difference arising from different optimization options would be mitigated~\cite{david2017similarity}. We utilize the built-in toolkit of LLVM~\cite{lattner2002llvm} to re-optimize the LLVM IR obtained after lifting the binary. 

\subsection{CPG}
Compilers generate a code's \acrfull{ast} by parsing the programming language, followed by semantic validation. The \acrshort{ast} serves as the fundamental structure for representing code at a low level and captures more semantic information than \acrshort{cfg}s~\cite{yang2021asteria}. \acrshort{cpg} is an enhanced version of the \acrshort{ast}. It explicitly incorporates information such as data dependencies and program dependencies, providing a richer semantic representation than the \acrshort{ast} alone. With the CPG in hand, we can infer the corresponding code's syntax and semantic information in a uniform graph representation. After decompiling the re-optimized LLVM IR into pseudo-C code, we use Joern~\cite{Yamaguchi_2014} to transform C code into CPG~\cite{Alex_Denisov_2021}.

\subsection{Graph Neural Network}
The process of lifting, code re-optimization, and \acrshort{cpg} reconstruction does not always produce identical graphs for similar functions. Some structural differences can still exist, and some information remains unrecoverable, leading to variations in the control flow of the codes. These differences are reflected in the graphs as structural discrepancies. Hence, we need a graph-matching algorithm that tolerates partial structural differences based on semantic relationships, enabling the determination of graph similarity between two semantically similar functions even when there are differences in their graph structures. As \acrshort{gnn}s comprehensively consider not only the graph topology but also global information based on the features of nodes and edges~\cite{velivckovic2017graph, li2019graph}, we incorporate semantic information into the node features of \acrshort{cpg}s and train a \acrshort{gnn} based framework capable of evaluating the similarity between two \acrshort{cpg}s.

\section{Evaluation}

\begin{table*}[tb]
  \centering
  \caption{The similarity score reported by BinDiff~\cite{Bindiff_2011} before and after applying lifting, re-optimization and re-compilation}
\scalebox{0.9}{
    \begin{tabular}{ccccc|ccc|ccc|ccc|ccc}
    \toprule
          &       & \multicolumn{15}{c}{O3} \\
\cmidrule{3-17}          &       & \multicolumn{3}{c}{ARM} & \multicolumn{3}{c}{PowerPC} & \multicolumn{3}{c}{MIPS} & \multicolumn{3}{c}{X86} & \multicolumn{3}{c}{AVG} \\
\cmidrule{3-17}          &       & Before & After & Inc   & Before & After & Inc   & Before & After & Inc   & Before & After & Inc   & Before & After & Inc \\
    \midrule
    \multicolumn{1}{c|}{\multirow{4}[2]{*}{O0}} & ARM   & 0.239 & 0.676 & 183\% & 0.415 & 0.657 & 58\%  & 0.346 & 0.672 & 94\%  & 0.386 & 0.628 & 63\%  & 0.347 & 0.658 & 90\% \\
    \multicolumn{1}{c|}{} & PowerPC   & 0.233 & 0.573 & 146\% & 0.373 & 0.622 & 67\%  & 0.348 & 0.578 & 66\%  & 0.376 & 0.555 & 48\%  & 0.333 & 0.582 & 75\% \\
    \multicolumn{1}{c|}{} & MIPS  & 0.258 & 0.591 & 129\% & 0.435 & 0.601 & 38\%  & 0.532 & 0.672 & 26\%  & 0.462 & 0.596 & 29\%  & 0.422 & 0.615 & 46\% \\
    \multicolumn{1}{c|}{} & X86   & 0.233 & 0.626 & 169\% & 0.415 & 0.641 & 54\%  & 0.347 & 0.697 & 101\% & 0.400 & 0.632 & 58\%  & 0.349 & 0.649 & 86\% \\
    \midrule
    \multicolumn{2}{c}{AVG} & 0.241 & 0.617 & 157\% & 0.410 & 0.630 & 54\%  & 0.393 & 0.655 & 72\%  & 0.406 & 0.603 & 49\%  & 0.362 & 0.626 & 74\% \\
    \bottomrule
    \end{tabular}%
}
  \label{tab:before_similar}%
\end{table*}%

In this section, we evalute~\name~on open-source software~\cite{alves2014openplc} and public datasets~\cite{marcelli2022machine}, we aim to answer the following evaluation questions (EQs): 
\begin{enumerate}
    \item \textbf{EQ 1:} Can lifters and optimizers reduce differences between binary code functions from different compilation environments? 
    \item \textbf{EQ 2:} Compared with state-of-the-art methods, how much speedup can \name~achieve while keeping similar performance?
\end{enumerate}

\subsection{BinDiff Enhancement}

To answer \textbf{EQ 1}, we choose to apply lifting and re-optimization on binary files to be compared among each other with the BinDiff~\cite{Bindiff_2011} tool, a widely used open-source framework that computes structural similarity scores between functions in binary executables. This enables us to show substantial improvement in the results. To do this, we carry out the following experiment using the OpenPLC dataset~\cite{alves2014openplc}. We take the default example ladder logic file from OpenPLC Editor~\cite{inicial_2022} as the input and compile it into binary files for four different architectures: MIPS, X86, PowerPC, and ARM. We use two extreme optimization levels: -O0 for no optimization and -O3 for the highest optimization. We then use the BinDiff~\cite{Bindiff_2011} and IDA Pro~\cite{Hex_1996} to have the similarity score for the same function but from binaries with different compilation environments. In the BinDiff, the similarity score ranges from 0 to 1.

To simulate conditions involving cross-architecture and cross-optimization option scenarios, a function that originates from -O0 of a certain architecture is compared with the same name functions from all architectures using the -O3. Then, we extract the corresponding similarity scores from Bindiff, and calculate the average similarity score from all the functions from that binary as the total result for one compilation environment versus another compilation environment. Next, we lift, re-optimize, and recompile the binary files using the x86-O3 setup. Subsequently, we repeat the same process to observe changes in the similarity scores.

The results are displayed in Table~\ref{tab:before_similar} to provide a clear visual representation. The percentage increase in the similarity score for each comparison is presented in the increase (Inc.) column. Furthermore, the average similarity scores for each column and row are calculated and presented in the last row and column, respectively. From the results, we conclude our answer to the \textbf{EQ 1}.

\begin{tcolorbox}[
    colback=gray!10, 
    colframe=black,  
    sharp corners=northwest, 
    rounded corners=all, 
    width=\linewidth, 
    boxrule=0.8pt, 
    arc=6pt, 
    left=2mm, 
    right=2mm, 
    ]
\textbf{Answer to EQ 1: } The similarity scores reported by BinDiff show a significant increase across all comparisons by 74\% on average, highlighting the effectiveness of our method in mitigating differences between optimization levels and distinct architectures.
\end{tcolorbox}

Since BinDiff relies on graph matching and some functions cannot be fully re-optimized, some differences remain between graphs. Despite major improvement, those differences prevent the similarity scores from reaching 1 with BinDiff. Moreover, we noticed that BinDiff performs very poorly when comparing stripped binaries with non-stripped ones. 

\subsection{Public Dataset Evaluation}
We evaluate \name~with the state-of-the-art models using the identical dataset proposed by~\cite{marcelli2022machine} to answer \textbf{EQ 2}. Our experimental platform is a general-purpose PC with an Intel i7-12700F CPU and an RTX 2080 Ti GPU. We align the exact GPU specifications and other experiment conditions as defined in~\cite{marcelli2022machine}. We separately train and evaluate our model on two distinct, non-overlapping datasets from~\cite{marcelli2022machine} to simulate scenarios where the model encounters unseen binary files. From~\cite{marcelli2022machine}, \texttt{Recall@1} is used to evaluate the accuracy of a model's first prediction, while \texttt{secs/100 functions} measures its inference speed. We run~\name~on CPU and GPU separately. Table~\ref{tab:speed} shows our result compared to other models in~\cite{marcelli2022machine}. Given that preprocessing, lifting, and re-optimization utilize external tools outside the scope of our control, we restrict our evaluation to model inference time, which is consistent with that used in~\cite{marcelli2022machine}. From the results, we conclude our answer to the \textbf{EQ 2}.

\begin{table}[tb]
\centering
\resizebox{\columnwidth}{!}{ 
\begin{tabular}{lcccccc}
\toprule
 & \makecell{ Zeek\\\scriptsize on GPU} & \makecell{Gemini\\\scriptsize on GPU} & \makecell{Asm2Vec\\\scriptsize on GPU} & \makecell{GMN\\\scriptsize on GPU} & \makecell{ReGraph\\\scriptsize on CPU} & \makecell{ ReGraph\\\scriptsize on GPU} \\
\midrule
\makecell{Recall@1} & 0.21 & 0.33 & 0.18 & 0.54 & \textbf{0.65} & \textbf{0.65} \\
\makecell{\shortstack{Time\\\tiny secs/100 functions}} & 0.091 & 0.135 & 7.17 & 0.876 & 0.283 & \textbf{0.011} \\
\bottomrule
\end{tabular}
} 
\caption{Perf. comparison with SotA methods from~\cite{marcelli2022machine}.}
\label{tab:speed}
\vspace{-8mm}
\end{table}

\begin{tcolorbox}[
    colback=gray!10, 
    colframe=black,  
    sharp corners=northwest, 
    rounded corners=all, 
    width=\linewidth, 
    boxrule=0.8pt, 
    arc=6pt, 
    left=2mm, 
    right=2mm, 
    ]
\textbf{Answer to EQ 2:} While maintaining high accuracy as measured by the Recall@1 metric, \name~achieves a speedup of over 700× compared to the NLP-based model Asm2Vec~\cite{ding2019asm2vec} and 9× compared to Zeek~\cite{zeek}. Moreover, it performs efficiently even when running solely on a CPU.
\end{tcolorbox}

\section{Conclusion and Future Work}

In this paper, we propose \name, a BCSD framework that effectively handles variations introduced by different compilation environments. \name~mitigates discrepancies caused by different compilation options by leveraging lifters and re-optimizers. With a lightweight GNN model, our experiments empirically demonstrate that \name~can yield promising results compared with those state-of-the-art solutions but with considerably less resource consumption. Furthermore, our experiment demonstrates that lifting and re-optimization can serve as a meta-method to enhance the performance of existing BCSD tools. We hope that our tool will inspire new perspectives in binary code analysis. In the future, we plan to extend our framework to the snippet level to enhance robustness.

\section*{Tool Availability}
We open-source \name~ and pre-trained models on GitHub: \textbf{https:
//github.com/damaoooo/ReGraph}. Additionally, our demo video can be seen on YouTube: \textbf{https://youtu.be/5CSJkZh89hs}.

\bibliographystyle{ACM-Reference-Format}
\bibliography{biblio}


\begin{thebibliography}{37}


\ifx \showCODEN    \undefined \def \showCODEN     #1{\unskip}     \fi
\ifx \showDOI      \undefined \def \showDOI       #1{#1}\fi
\ifx \showISBNx    \undefined \def \showISBNx     #1{\unskip}     \fi
\ifx \showISBNxiii \undefined \def \showISBNxiii  #1{\unskip}     \fi
\ifx \showISSN     \undefined \def \showISSN      #1{\unskip}     \fi
\ifx \showLCCN     \undefined \def \showLCCN      #1{\unskip}     \fi
\ifx \shownote     \undefined \def \shownote      #1{#1}          \fi
\ifx \showarticletitle \undefined \def \showarticletitle #1{#1}   \fi
\ifx \showURL      \undefined \def \showURL       {\relax}        \fi
\providecommand\bibfield[2]{#2}
\providecommand\bibinfo[2]{#2}
\providecommand\natexlab[1]{#1}
\providecommand\showeprint[2][]{arXiv:#2}

\bibitem[Agency(2019)]%
        {Agency_2019}
\bibfield{author}{\bibinfo{person}{NationalSecurity Agency}.} \bibinfo{year}{2019}\natexlab{}.
\newblock \bibinfo{title}{NationalSecurityAgency/Ghidra: Ghidra is a software reverse engineering (SRE) framework}.
\newblock
\newblock
\urldef\tempurl%
\url{https://github.com/NationalSecurityAgency/ghidra}
\showURL{%
\tempurl}


\bibitem[Alex~Denisov(2021)]%
        {Alex_Denisov_2021}
\bibfield{author}{\bibinfo{person}{Fabian~Yamaguchi Alex~Denisov}.} \bibinfo{year}{2021}\natexlab{}.
\newblock \bibinfo{title}{LLVM meets code property graphs}.
\newblock
\newblock
\urldef\tempurl%
\url{https://blog.llvm.org/posts/2021-02-23-llvm-meets-code-property-graphs/}
\showURL{%
\tempurl}


\bibitem[Alves et~al\mbox{.}(2014)]%
        {alves2014openplc}
\bibfield{author}{\bibinfo{person}{Thiago~Rodrigues Alves}, \bibinfo{person}{Mario Buratto}, \bibinfo{person}{Flavio~Mauricio De~Souza}, {and} \bibinfo{person}{Thelma~Virginia Rodrigues}.} \bibinfo{year}{2014}\natexlab{}.
\newblock \showarticletitle{OpenPLC: An open source alternative to automation}. In \bibinfo{booktitle}{\emph{IEEE Global Humanitarian Technology Conference (GHTC 2014)}}. IEEE, \bibinfo{pages}{585--589}.
\newblock


\bibitem[Basit and Jarzabek(2005)]%
        {basit2005detecting}
\bibfield{author}{\bibinfo{person}{Hamid~Abdul Basit} {and} \bibinfo{person}{Stan Jarzabek}.} \bibinfo{year}{2005}\natexlab{}.
\newblock \showarticletitle{Detecting higher-level similarity patterns in programs}.
\newblock \bibinfo{journal}{\emph{ACM Sigsoft Software engineering notes}} \bibinfo{volume}{30}, \bibinfo{number}{5} (\bibinfo{year}{2005}), \bibinfo{pages}{156--165}.
\newblock


\bibitem[Chandramohan et~al\mbox{.}(2016)]%
        {chandramohan2016bingo}
\bibfield{author}{\bibinfo{person}{Mahinthan Chandramohan}, \bibinfo{person}{Yinxing Xue}, \bibinfo{person}{Zhengzi Xu}, \bibinfo{person}{Yang Liu}, \bibinfo{person}{Chia~Yuan Cho}, {and} \bibinfo{person}{Hee Beng~Kuan Tan}.} \bibinfo{year}{2016}\natexlab{}.
\newblock \showarticletitle{Bingo: Cross-architecture cross-os binary search}. In \bibinfo{booktitle}{\emph{Proceedings of the 2016 24th ACM SIGSOFT International Symposium on Foundations of Software Engineering}}. \bibinfo{pages}{678--689}.
\newblock


\bibitem[Clang(2024)]%
        {Clang_2024}
\bibfield{author}{\bibinfo{person}{TI Clang}.} \bibinfo{year}{2024}\natexlab{}.
\newblock \bibinfo{title}{1.3.7. optimization options}.
\newblock
\newblock
\urldef\tempurl%
\url{https://software-dl.ti.com/codegen/docs/tiarmclang/compiler_tools_user_guide/compiler_manual/using_compiler/compiler_options/optimization_options.html}
\showURL{%
\tempurl}


\bibitem[David et~al\mbox{.}(2017)]%
        {david2017similarity}
\bibfield{author}{\bibinfo{person}{Yaniv David}, \bibinfo{person}{Nimrod Partush}, {and} \bibinfo{person}{Eran Yahav}.} \bibinfo{year}{2017}\natexlab{}.
\newblock \showarticletitle{Similarity of binaries through re-optimization}. In \bibinfo{booktitle}{\emph{Proceedings of the 38th ACM SIGPLAN conference on programming language design and implementation}}. \bibinfo{pages}{79--94}.
\newblock


\bibitem[David and Yahav(2014)]%
        {david2014tracelet}
\bibfield{author}{\bibinfo{person}{Yaniv David} {and} \bibinfo{person}{Eran Yahav}.} \bibinfo{year}{2014}\natexlab{}.
\newblock \showarticletitle{Tracelet-based code search in executables}.
\newblock \bibinfo{journal}{\emph{Acm Sigplan Notices}} \bibinfo{volume}{49}, \bibinfo{number}{6} (\bibinfo{year}{2014}), \bibinfo{pages}{349--360}.
\newblock


\bibitem[Ding et~al\mbox{.}(2019)]%
        {ding2019asm2vec}
\bibfield{author}{\bibinfo{person}{Steven~HH Ding}, \bibinfo{person}{Benjamin~CM Fung}, {and} \bibinfo{person}{Philippe Charland}.} \bibinfo{year}{2019}\natexlab{}.
\newblock \showarticletitle{Asm2vec: Boosting static representation robustness for binary clone search against code obfuscation and compiler optimization}. In \bibinfo{booktitle}{\emph{2019 IEEE Symposium on Security and Privacy (SP)}}. IEEE, \bibinfo{pages}{472--489}.
\newblock


\bibitem[GmbH(2011)]%
        {Bindiff_2011}
\bibfield{author}{\bibinfo{person}{GmbH}.} \bibinfo{year}{2011}\natexlab{}.
\newblock
\newblock
\urldef\tempurl%
\url{https://www.zynamics.com/bindiff.html}
\showURL{%
\tempurl}


\bibitem[GNU({[n.\,d.]})]%
        {GNU_strip}
\bibfield{author}{\bibinfo{person}{GNU}.} \bibinfo{year}{[n.\,d.]}\natexlab{}.
\newblock
\newblock
\urldef\tempurl%
\url{https://ftp.gnu.org/old-gnu/Manuals/binutils-2.12/html_node/binutils_10.html}
\showURL{%
\tempurl}


\bibitem[GNU(2023)]%
        {GNU_2023}
\bibfield{author}{\bibinfo{person}{GCC GNU}.} \bibinfo{year}{2023}\natexlab{}.
\newblock \bibinfo{title}{Options That Control Optimization}.
\newblock
\newblock
\urldef\tempurl%
\url{https://gcc.gnu.org/onlinedocs/gcc/Optimize-Options.html}
\showURL{%
\tempurl}


\bibitem[Hex(1996)]%
        {Hex_1996}
\bibfield{author}{\bibinfo{person}{Rays Hex}.} \bibinfo{year}{1996}\natexlab{}.
\newblock \bibinfo{title}{Ida Pro}.
\newblock
\newblock
\urldef\tempurl%
\url{https://hex-rays.com/ida-pro/}
\showURL{%
\tempurl}


\bibitem[Hu et~al\mbox{.}(2016)]%
        {hu2016cross}
\bibfield{author}{\bibinfo{person}{Yikun Hu}, \bibinfo{person}{Yuanyuan Zhang}, \bibinfo{person}{Juanru Li}, {and} \bibinfo{person}{Dawu Gu}.} \bibinfo{year}{2016}\natexlab{}.
\newblock \showarticletitle{Cross-architecture binary semantics understanding via similar code comparison}. In \bibinfo{booktitle}{\emph{2016 IEEE 23rd international conference on software analysis, evolution, and reengineering (SANER)}}, Vol.~\bibinfo{volume}{1}. IEEE, \bibinfo{pages}{57--67}.
\newblock


\bibitem[Huang et~al\mbox{.}(2017)]%
        {huang2017binsequence}
\bibfield{author}{\bibinfo{person}{He Huang}, \bibinfo{person}{Amr~M Youssef}, {and} \bibinfo{person}{Mourad Debbabi}.} \bibinfo{year}{2017}\natexlab{}.
\newblock \showarticletitle{Binsequence: Fast, accurate and scalable binary code reuse detection}. In \bibinfo{booktitle}{\emph{Proceedings of the 2017 ACM on Asia conference on computer and communications security}}. \bibinfo{pages}{155--166}.
\newblock


\bibitem[inicial(2022)]%
        {inicial_2022}
\bibfield{author}{\bibinfo{person}{Página inicial}.} \bibinfo{year}{2022}\natexlab{}.
\newblock
\newblock
\urldef\tempurl%
\url{https://openplcproject.com/docs/3-1-openplc-editor-overview/}
\showURL{%
\tempurl}


\bibitem[K{\v{r}}oustek et~al\mbox{.}(2017)]%
        {kvroustek2017retdec}
\bibfield{author}{\bibinfo{person}{Jakub K{\v{r}}oustek}, \bibinfo{person}{Peter Matula}, {and} \bibinfo{person}{P Zemek}.} \bibinfo{year}{2017}\natexlab{}.
\newblock \showarticletitle{Retdec: An open-source machine-code decompiler}. In \bibinfo{booktitle}{\emph{July 2018}}.
\newblock


\bibitem[Lattner(2002)]%
        {lattner2002llvm}
\bibfield{author}{\bibinfo{person}{Chris~Arthur Lattner}.} \bibinfo{year}{2002}\natexlab{}.
\newblock \showarticletitle{LLVM: An infrastructure for multi-stage optimization}.
\newblock  (\bibinfo{year}{2002}).
\newblock


\bibitem[Li et~al\mbox{.}(2019)]%
        {li2019graph}
\bibfield{author}{\bibinfo{person}{Yujia Li}, \bibinfo{person}{Chenjie Gu}, \bibinfo{person}{Thomas Dullien}, \bibinfo{person}{Oriol Vinyals}, {and} \bibinfo{person}{Pushmeet Kohli}.} \bibinfo{year}{2019}\natexlab{}.
\newblock \showarticletitle{Graph matching networks for learning the similarity of graph structured objects}. In \bibinfo{booktitle}{\emph{International conference on machine learning}}. PMLR, \bibinfo{pages}{3835--3845}.
\newblock


\bibitem[Luo et~al\mbox{.}(2014)]%
        {luo2014semantics}
\bibfield{author}{\bibinfo{person}{Lannan Luo}, \bibinfo{person}{Jiang Ming}, \bibinfo{person}{Dinghao Wu}, \bibinfo{person}{Peng Liu}, {and} \bibinfo{person}{Sencun Zhu}.} \bibinfo{year}{2014}\natexlab{}.
\newblock \showarticletitle{Semantics-based obfuscation-resilient binary code similarity comparison with applications to software plagiarism detection}. In \bibinfo{booktitle}{\emph{Proceedings of the 22nd ACM SIGSOFT international symposium on foundations of software engineering}}. \bibinfo{pages}{389--400}.
\newblock


\bibitem[Luo et~al\mbox{.}(2023)]%
        {luo2023vulhawk}
\bibfield{author}{\bibinfo{person}{Zhenhao Luo}, \bibinfo{person}{Pengfei Wang}, \bibinfo{person}{Baosheng Wang}, \bibinfo{person}{Yong Tang}, \bibinfo{person}{Wei Xie}, \bibinfo{person}{Xu Zhou}, \bibinfo{person}{Danjun Liu}, {and} \bibinfo{person}{Kai Lu}.} \bibinfo{year}{2023}\natexlab{}.
\newblock \showarticletitle{VulHawk: Cross-architecture Vulnerability Detection with Entropy-based Binary Code Search.}. In \bibinfo{booktitle}{\emph{NDSS}}.
\newblock


\bibitem[Marcelli et~al\mbox{.}(2022)]%
        {marcelli2022machine}
\bibfield{author}{\bibinfo{person}{Andrea Marcelli}, \bibinfo{person}{Mariano Graziano}, \bibinfo{person}{Xabier Ugarte-Pedrero}, \bibinfo{person}{Yanick Fratantonio}, \bibinfo{person}{Mohamad Mansouri}, {and} \bibinfo{person}{Davide Balzarotti}.} \bibinfo{year}{2022}\natexlab{}.
\newblock \showarticletitle{How machine learning is solving the binary function similarity problem}. In \bibinfo{booktitle}{\emph{31st USENIX Security Symposium (USENIX Security 22)}}. \bibinfo{pages}{2099--2116}.
\newblock


\bibitem[Mengin and Rossi(2021)]%
        {mengin2021binary}
\bibfield{author}{\bibinfo{person}{Elie Mengin} {and} \bibinfo{person}{Fabrice Rossi}.} \bibinfo{year}{2021}\natexlab{}.
\newblock \showarticletitle{Binary diffing as a network alignment problem via belief propagation}. In \bibinfo{booktitle}{\emph{2021 36th IEEE/ACM International Conference on Automated Software Engineering (ASE)}}. IEEE, \bibinfo{pages}{967--978}.
\newblock


\bibitem[Pardoe et~al\mbox{.}(2018)]%
        {stat462_correlation}
\bibfield{author}{\bibinfo{person}{Iain Pardoe}, \bibinfo{person}{Laura Simon}, {and} \bibinfo{person}{Derek Young}.} \bibinfo{year}{2018}\natexlab{}.
\newblock \bibinfo{title}{2.6 - (Pearson) Correlation Coefficient r}.
\newblock
\newblock
\urldef\tempurl%
\url{https://online.stat.psu.edu/stat462/node/96/}
\showURL{%
\tempurl}
\newblock
\shownote{Accessed: 2025-03-11}.


\bibitem[Pewny et~al\mbox{.}(2015)]%
        {pewny2015cross}
\bibfield{author}{\bibinfo{person}{Jannik Pewny}, \bibinfo{person}{Behrad Garmany}, \bibinfo{person}{Robert Gawlik}, \bibinfo{person}{Christian Rossow}, {and} \bibinfo{person}{Thorsten Holz}.} \bibinfo{year}{2015}\natexlab{}.
\newblock \showarticletitle{Cross-architecture bug search in binary executables}. In \bibinfo{booktitle}{\emph{2015 IEEE Symposium on Security and Privacy}}. IEEE, \bibinfo{pages}{709--724}.
\newblock


\bibitem[Shalev and Partush(2018)]%
        {zeek}
\bibfield{author}{\bibinfo{person}{Noam Shalev} {and} \bibinfo{person}{Nimrod Partush}.} \bibinfo{year}{2018}\natexlab{}.
\newblock \showarticletitle{Binary similarity detection using machine learning}. In \bibinfo{booktitle}{\emph{Proceedings of the 13th Workshop on Programming Languages and Analysis for Security}}. \bibinfo{pages}{42--47}.
\newblock


\bibitem[Shirani et~al\mbox{.}(2018)]%
        {shirani2018b}
\bibfield{author}{\bibinfo{person}{Paria Shirani}, \bibinfo{person}{Leo Collard}, \bibinfo{person}{Basile~L Agba}, \bibinfo{person}{Bernard Lebel}, \bibinfo{person}{Mourad Debbabi}, \bibinfo{person}{Lingyu Wang}, {and} \bibinfo{person}{Aiman Hanna}.} \bibinfo{year}{2018}\natexlab{}.
\newblock \showarticletitle{B in a rm: Scalable and efficient detection of vulnerabilities in firmware images of intelligent electronic devices}. In \bibinfo{booktitle}{\emph{Detection of Intrusions and Malware, and Vulnerability Assessment: 15th International Conference, DIMVA 2018, Saclay, France, June 28--29, 2018, Proceedings 15}}. Springer, \bibinfo{pages}{114--138}.
\newblock


\bibitem[Veli{\v{c}}kovi{\'c} et~al\mbox{.}(2017)]%
        {velivckovic2017graph}
\bibfield{author}{\bibinfo{person}{Petar Veli{\v{c}}kovi{\'c}}, \bibinfo{person}{Guillem Cucurull}, \bibinfo{person}{Arantxa Casanova}, \bibinfo{person}{Adriana Romero}, \bibinfo{person}{Pietro Lio}, {and} \bibinfo{person}{Yoshua Bengio}.} \bibinfo{year}{2017}\natexlab{}.
\newblock \showarticletitle{Graph attention networks}.
\newblock \bibinfo{journal}{\emph{arXiv preprint arXiv:1710.10903}} (\bibinfo{year}{2017}).
\newblock


\bibitem[Wang et~al\mbox{.}(2024)]%
        {wang2024binenhance}
\bibfield{author}{\bibinfo{person}{Yongpan Wang}, \bibinfo{person}{Hong Li}, \bibinfo{person}{Xiaojie Zhu}, \bibinfo{person}{Siyuan Li}, \bibinfo{person}{Chaopeng Dong}, \bibinfo{person}{Shouguo Yang}, {and} \bibinfo{person}{Kangyuan Qin}.} \bibinfo{year}{2024}\natexlab{}.
\newblock \showarticletitle{BinEnhance: An Enhancement Framework Based on External Environment Semantics for Binary Code Search}.
\newblock \bibinfo{journal}{\emph{arXiv preprint arXiv:2411.01102}} (\bibinfo{year}{2024}).
\newblock


\bibitem[xorpd({[n.\,d.]})]%
        {xorpd}
\bibfield{author}{\bibinfo{person}{xorpd}.} \bibinfo{year}{[n.\,d.]}\natexlab{}.
\newblock \bibinfo{title}{FCATALOG}.
\newblock
\newblock
\urldef\tempurl%
\url{https://www.xorpd.net/pages/fcatalog.html}
\showURL{%
\tempurl}


\bibitem[Xu et~al\mbox{.}(2017)]%
        {xu2017neural}
\bibfield{author}{\bibinfo{person}{Xiaojun Xu}, \bibinfo{person}{Chang Liu}, \bibinfo{person}{Qian Feng}, \bibinfo{person}{Heng Yin}, \bibinfo{person}{Le Song}, {and} \bibinfo{person}{Dawn Song}.} \bibinfo{year}{2017}\natexlab{}.
\newblock \showarticletitle{Neural network-based graph embedding for cross-platform binary code similarity detection}. In \bibinfo{booktitle}{\emph{Proceedings of the 2017 ACM SIGSAC conference on computer and communications security}}. \bibinfo{pages}{363--376}.
\newblock


\bibitem[Xu et~al\mbox{.}(2021)]%
        {xu2021innovative}
\bibfield{author}{\bibinfo{person}{Zimu Xu}, \bibinfo{person}{Maria~H Gonzalez-Serrano}, \bibinfo{person}{Rocco Porreca}, {and} \bibinfo{person}{Paul Jones}.} \bibinfo{year}{2021}\natexlab{}.
\newblock \showarticletitle{Innovative sports-embedded gambling promotion: A study of spectators’ enjoyment and gambling intention during XFL games}.
\newblock \bibinfo{journal}{\emph{Journal of Business Research}}  \bibinfo{volume}{131} (\bibinfo{year}{2021}), \bibinfo{pages}{206--216}.
\newblock


\bibitem[Yamaguchi(2014)]%
        {Yamaguchi_2014}
\bibfield{author}{\bibinfo{person}{Yamaguchi}.} \bibinfo{year}{2014}\natexlab{}.
\newblock \bibinfo{title}{The bug hunter’s workbench}.
\newblock
\newblock
\urldef\tempurl%
\url{https://joern.io/}
\showURL{%
\tempurl}


\bibitem[Yang et~al\mbox{.}(2021)]%
        {yang2021asteria}
\bibfield{author}{\bibinfo{person}{Shouguo Yang}, \bibinfo{person}{Long Cheng}, \bibinfo{person}{Yicheng Zeng}, \bibinfo{person}{Zhe Lang}, \bibinfo{person}{Hongsong Zhu}, {and} \bibinfo{person}{Zhiqiang Shi}.} \bibinfo{year}{2021}\natexlab{}.
\newblock \showarticletitle{Asteria: Deep learning-based AST-encoding for cross-platform binary code similarity detection}. In \bibinfo{booktitle}{\emph{2021 51st Annual IEEE/IFIP International Conference on Dependable Systems and Networks (DSN)}}. IEEE, \bibinfo{pages}{224--236}.
\newblock


\bibitem[Yu et~al\mbox{.}(2020a)]%
        {yu2020order}
\bibfield{author}{\bibinfo{person}{Zeping Yu}, \bibinfo{person}{Rui Cao}, \bibinfo{person}{Qiyi Tang}, \bibinfo{person}{Sen Nie}, \bibinfo{person}{Junzhou Huang}, {and} \bibinfo{person}{Shi Wu}.} \bibinfo{year}{2020}\natexlab{a}.
\newblock \showarticletitle{Order matters: Semantic-aware neural networks for binary code similarity detection}. In \bibinfo{booktitle}{\emph{Proceedings of the AAAI conference on artificial intelligence}}, Vol.~\bibinfo{volume}{34}. \bibinfo{pages}{1145--1152}.
\newblock


\bibitem[Yu et~al\mbox{.}(2020b)]%
        {yu2020codecmr}
\bibfield{author}{\bibinfo{person}{Zeping Yu}, \bibinfo{person}{Wenxin Zheng}, \bibinfo{person}{Jiaqi Wang}, \bibinfo{person}{Qiyi Tang}, \bibinfo{person}{Sen Nie}, {and} \bibinfo{person}{Shi Wu}.} \bibinfo{year}{2020}\natexlab{b}.
\newblock \showarticletitle{Codecmr: Cross-modal retrieval for function-level binary source code matching}.
\newblock \bibinfo{journal}{\emph{Advances in Neural Information Processing Systems}}  \bibinfo{volume}{33} (\bibinfo{year}{2020}), \bibinfo{pages}{3872--3883}.
\newblock


\bibitem[Zhao et~al\mbox{.}(2022)]%
        {zhao2022large}
\bibfield{author}{\bibinfo{person}{Binbin Zhao}, \bibinfo{person}{Shouling Ji}, \bibinfo{person}{Jiacheng Xu}, \bibinfo{person}{Yuan Tian}, \bibinfo{person}{Qiuyang Wei}, \bibinfo{person}{Qinying Wang}, \bibinfo{person}{Chenyang Lyu}, \bibinfo{person}{Xuhong Zhang}, \bibinfo{person}{Changting Lin}, \bibinfo{person}{Jingzheng Wu}, {et~al\mbox{.}}} \bibinfo{year}{2022}\natexlab{}.
\newblock \showarticletitle{A large-scale empirical analysis of the vulnerabilities introduced by third-party components in IoT firmware}. In \bibinfo{booktitle}{\emph{Proceedings of the 31st ACM SIGSOFT International Symposium on Software Testing and Analysis}}. \bibinfo{pages}{442--454}.
\newblock


\end{thebibliography}

\end{document}